# Heat capacities of elastic solids

József Garai
Independent Researcher
E-mail: jozsef.garai@fiu.edu
Ph: 786-247-5414

**Abstract**
   The work function is embedded in the equation describing the relationship between the constant volume and constant pressure heat capacities. The modification of the work function results that the relationship between these quantities must be changed accordingly. Using the newly derived work functions of elastic solids the description of the heat capacities and the relationship between the heat capacities are given for solid phase.

The molar heat capacity $[c(\varphi)]_\gamma$ represents the energy required to raise the temperature of one mol substance by one degree of Kelvin. The general description can be given as:

$$[c(\varphi)]_\gamma = \frac{1}{n\delta T}\int_T^{T+\delta T}[\delta q(\varphi)]_\gamma = \frac{1}{n}\left[\frac{\partial q(\varphi)}{\partial T}\right]_\gamma \qquad (1)$$

where $\varphi = g(gas), \; s(solid), \; l(liquid)$, and $\gamma = p \;(pressure), \; V \;(volume)$, n is the number of moles, T is the temperature, and q is the heat supplied to the system. The energy added to the system can be utilized as thermal energy or mechanical work. The differential of the heat supplied to the system comprises both the thermal and work related energies. The thermodynamic conditions affects only the work related heat while the thermal related part remains unchanged regardless of the conditions.

$$[\delta q(\varphi)]_\gamma = \delta q(\varphi)^{thermal} + [\delta q(\varphi)^{work}]_\gamma \qquad (2)$$

The heat capacity in the same manner can be divided into thermal and work related parts. The thermal related part will be called molar thermal heat capacity $c(\varphi)^{thermal}$

$$c(\varphi)^{thermal} = \frac{1}{n\delta T}\int_T^{T+\delta T}\delta q(\varphi)^{thermal} = \frac{1}{n}\frac{\partial q(\varphi)^{thermal}}{\partial T} \qquad (3)$$

while the work related part will be called to molar work heat capacity $[c(\varphi)^{work}]_\gamma$

$$[c(\varphi)^{work}]_\gamma = \frac{1}{n\delta T}\int_T^{T+\delta T}[\delta q(\varphi)^{work}]_\gamma = \frac{1}{n}\left[\frac{\partial q(\varphi)^{work}}{\partial T}\right]_\gamma, \qquad (4)$$

The heat capacity is then the sum of the thermal and work related heat capacities.

$$[c(\varphi)]_\gamma = c(\varphi)^{\text{thermal}} + [c(\varphi)^{\text{work}}]_\gamma \tag{5}$$

In gas phase the molar heat capacity at constant volume is

$$[c(g)]_V = c(g)^{\text{thermal}} + [c(g)^{\text{work}}]_V = \frac{1}{n\delta T}\int_T^{T+\delta T}[\delta q(g)]_V \tag{6}$$

At constant volume no mechanical work is done and the molar work heat capacity is zero.

$$[c(g)^{\text{work}}]_V = 0 \tag{7}$$

The energy supplied to the system is entirely utilized as thermal energy.

$$c(g)^{\text{thermal}} = [c(g)]_V = \frac{1}{n\delta T}\int_T^{T+\delta T}[\delta q(g)]_V \tag{8}$$

According to the first law of thermodynamic the differential of the internal energy of a system [U] is equal with the sum of the differentials of the heat and work supplied to the system.

$$dU = \delta q + \delta w = nc_V dT - pdV \tag{9}$$

In gas phase at constant volume equation (9) reduces to

$$dU(g) = \delta q(g)_V = nc_V dT \tag{10}$$

At constant pressure the differential of the internal energy is:

$$dU(g) = [\delta q(g)]_p + [\delta w(g)]_p. \tag{11}$$

Combining equation (10) with (11) gives the differential of the heat supplied to the system at constant pressure.

$$[\delta q(g)]_p = [\delta q(g)]_V - [\delta w(g)]_p \tag{12}$$

The heat capacity at constant pressure is then

$$[c(g)]_p = \frac{1}{n\delta T}\int_T^{T+\delta T}(\delta q)_p = \frac{1}{n\delta T}\int_T^{T+\delta T}[\delta q(g)]_V + \frac{1}{n\delta T}\int_T^{T+\delta T}-[\delta w(g)]_p. \tag{13}$$

The first term in equation (13) is the molar heat capacity at constant volume [Eq. (8)] while the second term is the molar work heat capacity at constant pressure

$$[c(g)^{\text{work}}]_p = \frac{1}{n\delta T}\int_T^{T+\delta T}-[\delta w(g)]_p = \frac{1}{n\delta T}\int_T^{T+\delta T}p(dV)_p. \tag{14}$$

The relationship between the constant pressure and volume heat capacities in gas phase can be written as :

$$[c(g)]_p = [c(g)]_V + [c(g)^{\text{work}}]_p \tag{15}$$

Using the equation of state for ideal gasses

$$pV = nRT \Rightarrow p(dV)_p = nRdT \tag{16}$$

and substituting this result into equation (14) gives the molar work heat capacity for gasses

$$\left[c(g)^{work}\right]_p = \frac{1}{n\delta T}\int_T^{T+\delta T} p(dV)_p = \frac{1}{\delta T}\int_T^{T+\delta T} RdT = R = N_A k_B \qquad (17)$$

where R is the universal gas constant, $N_A$ is the Avogadro number, and $k_B$ is the Boltzmann constant. The relationship between the molar heat capacities at constant pressure and volume is

$$[c(g)]_p = [c(g)]_V + R. \qquad (18)$$

The current consensus in solid thermodynamics assumes that equation (8) is valid regardless of the phase; therefore, it can be used in the same form for solids[1].

$$[c(s)]_V = c(s)^{thermal} \approx c(s)^{Debye} \qquad (19)$$

where $c(s)^{Debye}$ is the theoretical value of the thermal heat capacity calculated by using the Debye concept. The relationship between the heat capacities is described[2] as :

$$[c(s)]_p = [c(s)]_V + \frac{VTK_T \alpha_{V_p}^2}{n} \qquad (20)$$

where $\alpha_{V_p}$ is the volume coefficient of thermal expansion and $K_T$ is the bulk modulus. In this study the thermodynamic description of the heat capacities and the relationship between the heat capacities will be investigated in detail in solid phase.

Investigating solids in a state of equilibrium it has been demonstrated that correlation between the temperature and the pressure can exist only at constant volume[3]. This correlation is irreversible and works in the temperature pressure direction. The limited communication between the temperature and the pressure put constraints on the conversion of the thermal and the mechanical energies. In solid phase the pressure correlates to the elastic related volume change while the temperature to the thermal related volume change. These volume changes are not compatible with the exception of constant volume condition when the expanded volume converts completely to compressed volume. Based on theoretical consideration it had been suggested that the thermal related volume change do not result mechanical work. Separating the thermal and elastic related volume changes the work functions for each of the thermodynamic conditions have been derived[4].

Since only the elastic volume change results work in solid phase the molar work heat capacity should be written as :

$$\left[c(s)^{work}\right]_\gamma = \frac{1}{n\delta T}\int_T^{T+\delta T} [\delta w(s)_T]_\gamma = \frac{1}{n\delta T}\int_T^{T+\delta T} -p(dV_T^{elastic})_\gamma. \qquad (21)$$

In solid phase contrarily to gasses the heat added to system does work on the system. The signs given in Eq. (14) for the differential of the work and for the work function have been changed in Eq. (21) accordingly. The work functions of solids[4] for constant pressure and volume are

$$[\Delta w(s)_T]_p = V_0 e^{\int_{T=0}^{T_i} \alpha_{V_p} dT} \left[ K_T - (K_T + p) e^{-\frac{p}{K_T}} \right] \left( e^{\int_{T_i}^{T_f} \alpha_{V_p} dT} - 1 \right) \quad (22)$$

$$[\Delta w(s)_T]_V = K_T V_0 e^{\alpha_{V_p} T_i} \left[ e^{\int_{T_i}^{T_f} \alpha_{V_p} dT} - 1 - \alpha_{V_p} e^{-\frac{p_i}{K_T}} (T_f - T_i) \right]. \quad (23)$$

Subscripts i and f are used for initial and final conditions respectively. Assuming that $T_i = T$ and that $T_f = T + 1K$ the internal work related to one degree of temperature change can be calculated

$$[\Delta w(s)_{\Delta T=1^0}]_p = V_0 e^{\alpha_{V_p} T} \left[ K_T - (K_T + p) e^{-\frac{p}{K_T}} \right] \left( e^{\alpha_{V_p}} - 1 \right) \quad (24)$$

and

$$[\Delta w(s)_{\Delta T=1^0}]_V = K_T V_0 e^{\alpha_{V_p} T} \left[ e^{\alpha_{V_p}} - 1 - \alpha_{V_p} e^{-\frac{p}{K_T}} \right]. \quad (25)$$

The molar work heat capacities for solids are

$$[c(s)^{work}]_p = \frac{1}{n} [\Delta w(s)_{\Delta T=1^0}]_p = \frac{1}{n} V_0 e^{\alpha_{V_p} T} \left[ K_T - (K_T + p) e^{-\frac{p}{K_T}} \right] \left( e^{\alpha_{V_p}} - 1 \right) \quad (26)$$

and

$$[c(s)^{work}]_V = \frac{1}{n} [\Delta w(s)_{\Delta T=1^0}]_V = \frac{1}{n} K_T V_0 e^{\alpha_{V_p} T} \left[ e^{\alpha_{V_p}} - 1 - \alpha_{V_p} e^{-\frac{p}{K_T}} \right]. \quad (27)$$

Substituting the molar volume at zero pressure $[(V_T^m)_{p=0}]$

$$(V_T^m)_{p=0} = \frac{1}{n} V_0 e^{\alpha_{V_p} T} \quad (28)$$

the heat capacities are at constant pressure $[c(s)]_p$

$$[c(s)]_p = [c(s)]^{thermal} + (V_T^m)_{p=0} \left[ K_T - (K_T + p) e^{-\frac{p}{K_T}} \right] \left( e^{\alpha_{V_p}} - 1 \right) \quad (29)$$

and at constant volume $[c(s)]_V$:

$$[c(s)]_V = [c(s)]^{\text{thermal}} + (V_T^m)_{p=0} K_T \left[ e^{\alpha_{V_p}} - 1 - \alpha_{V_p} e^{-\frac{p}{K_T}} \right]. \tag{30}$$

Combining equation (29) and (30) the relationship between the constant pressure and volume heat capacities is

$$[c(s)]_V - [c(s)]_p = (V_T^m)_{p=0} \left\{ K_T \left[ e^{\alpha_{V_p}} - 1 - \alpha_{V_p} e^{-\frac{p}{K_T}} \right] - \left[ K_T - (K_T + p) e^{-\frac{p}{K_T}} \right] \left( e^{\alpha_{V_p}} - 1 \right) \right\} \tag{31}$$

Simplifying equation (31), the relationship between constant pressure and constant volume molar heat capacities can be rewritten like:

$$[c(s)]_V - [c(s)]_p = V^m \left[ (K_T + p)\left(e^{\alpha_{V_p}} - 1\right) - K_T \alpha_{V_p} \right] \tag{32}$$

where

$$V^m = V_{p,T}^m = (V_T^m)_{p=0} e^{-\frac{p}{K_T}} = \frac{1}{n} V_0 e^{\alpha_{V_p} T} e^{-\frac{p}{K_T}}. \tag{33}$$

Assuming that $\alpha_{V_p} \times 1^0$ Kelvin is small compared to 1 then equation (32) can be written as:

$$[c(s)]_V - [c(s)]_p \approx p V^m \alpha_{V_p}. \tag{34}$$

It can be seen from equations (26) and (27) that in solid phase the mechanical work related to one degree of temperature change at constant volume is greater than at constant pressure.

$$\frac{1}{n\delta T} \int_T^{T+\delta T} [\delta w(s)]_V > \frac{1}{n\delta T} \int_T^{T+\delta T} [\delta w(s)]_p \tag{35}$$

The higher value of the mechanical work at constant volume indicates that the value of the heat capacity at constant volume should be higher than at constant pressure.

$$[c(s)]_V > [c(s)]_p > [c(s)]^{\text{thermal}} = [c(s)]_{p=0} \tag{36}$$

The temperature and pressure dependence of the molar work heat capacities and their relationship to each other is shown on Fig. 1.

Using the conventional approach [Eq. (20)] and calculating the molar volume heat capacities from experiments the theoretical and experimental values did not show good agreement for different minerals[5]. The inconsistencies might be resulting from the incorrect theoretical description of the heat capacity.

**Acknowledgement:** I would like to thank Alexandre Laugier for his helpful comments on the manuscript.

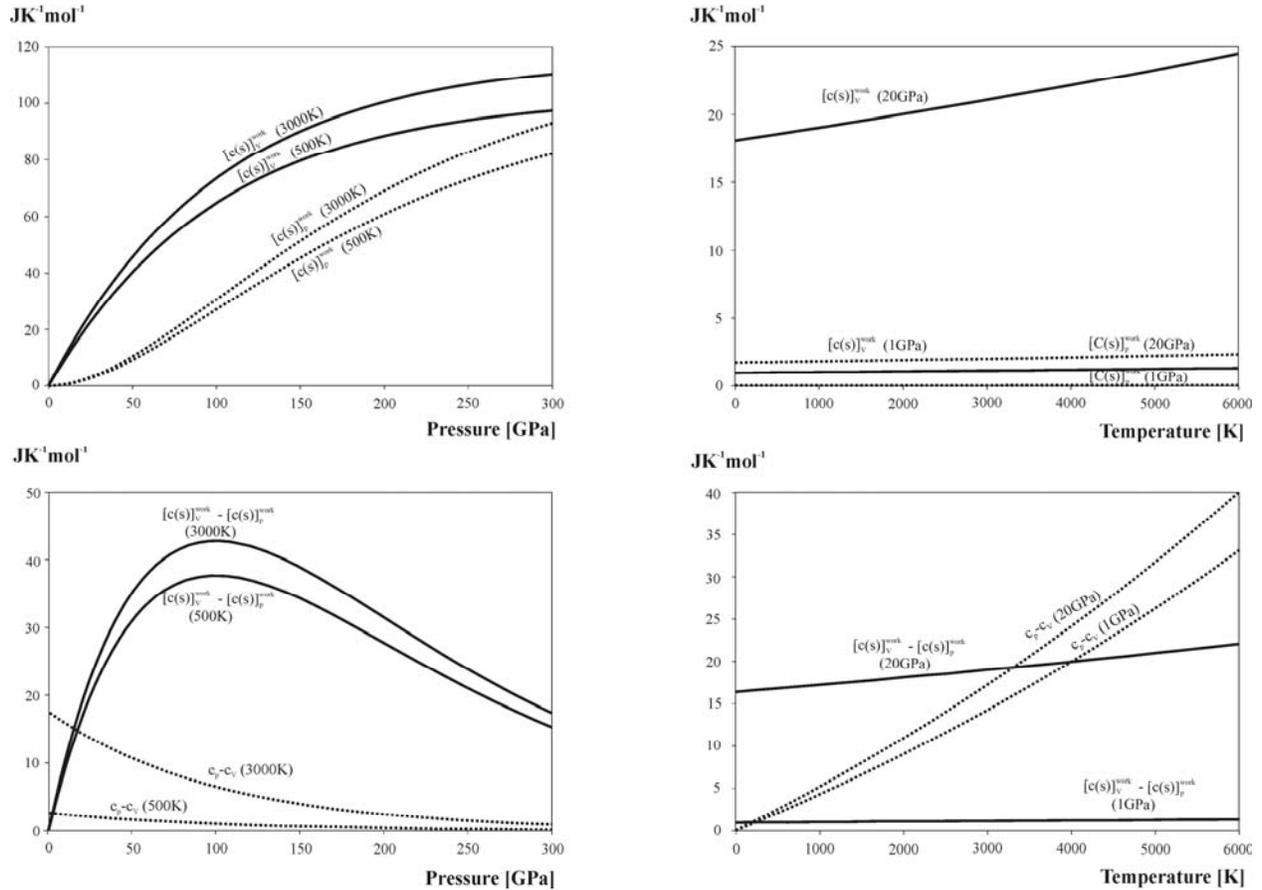

**FIG. 1**. The characteristic behavior of the molar work heat capacities is shown as function of pressure and temperature. The values $[c_p - c_v]$ calculated by the conventional method [Eq. (20)] are also plotted for comparison. The molar work heat capacities were calculated by using Eq. (29) and (30). It was assumed that the bulk modulus and the volume coefficient of thermal expansion are constant, and that $K_T = 1*10^{11} \text{Pa}$, $\alpha_{V_p} = 1*10^{-5} \text{K}^{-1}$, and $V_0^m = \left(V_{T=0}^m\right)_{p=0} = 2*10^{-5} \text{m}^3$.